# Full duplex communication using visible light

Yongchao Yang, Bingcheng Zhu, Yuanhang Li, Xumin Gao, Jialei Yuan, Hongbo Zhu and Yongjin Wang*

*Grünberg Research Centre, Nanjing University of Posts and Telecommunications, Nanjing 210003, China*

*Corresponding author: *wangyj@njupt.edu.cn*

**Abstract**

In this work, we propose, fabricate and characterize a full duplex communication system using visible light on a single chip. Both the suspended p-n junction InGaN/GaN multiple quantum well (MQW) devices and the suspended waveguides are obtained on a GaN-on-silicon platform by wafer-level processing. Two suspended p-n junction InGaN/GaN MQW devices that can both emit and detect light simultaneously are connected using suspended waveguides to form an in-plane visible light communication (VLC) system. The light that is emitted from one suspended p-n junction InGaN/GaN MQW device can induce a current in the device located at the other end of the waveguide via in-plane light coupling, thus leading to full duplex communication using visible light. This proof-of-concept in-plane VLC system paves the way towards the implementation of a full duplex communications system operating at the same frequency using visible light on a single chip.

In a typical visible light communication (VLC) system, a light-emitting diode (LED) serves as a transmitter that emits modulated light and a photodiode is used as a receiver to detect the light and complete the photon-electron conversion [1–3]. Previous experimental results show that existing photoelectric devices can only form a one-way VLC system in which data are transmitted from the LED to the photodiode using modulated light on the same channel. Two separate channels, i.e., one for uplink and one for downlink [4], are thus required to achieve full duplex communication. A p-n junction InGaN/GaN multiple quantum well (MQW) device has selectable light emission and photodetection functionalities [5–8]. Such a device, when acting as a receiver, is sensitive for detection of incoming light. However, in previous LED-to-LED VLC systems, bidirectional communication was obtained by temporal separation of the transmitter and receiver signals [9]. In principle, a single p-n junction InGaN/GaN MQW device can detect incoming light and convert the photons into an electrical current when operating in either LED mode or photodiode mode. This implies that such a device can realize light emission and photodetection simultaneously. This is promising for the development of full duplex VLC systems, in which a single channel can be used to double the spectral efficiency [10].

Here, we propose, fabricate and characterize a full duplex communication system using visible light on a single chip. Suspended p-n junction InGaN/GaN MQW devices act as both communication port and waveguide when the communication channels are fabricated on a GaN-on-silicon platform using a wafer-level process [11, 12]. Two suspended p-n junction InGaN/GaN MQW devices that can emit and detect light simultaneously are connected using suspended waveguides to form an in-plane VLC system. The light emitted from one communication port of this VLC system can induce electrical current signals in another communication port located at the other end of the system via in-plane light coupling. Full duplex communication is experimentally demonstrated on an in-plane VLC system that transmits and receives modulated light simultaneously on the same channel. The proposed device doubles the spectral efficiency, reduces the number of devices required for full-duplex VLC communications by half, and compresses the entire communication system down to the microscale.

The in-plane VLC system is implemented on a two-inch GaN-on-silicon wafer, and is based on a p-n junction InGaN/GaN MQW structure [13, 14]. Because the p-n junction InGaN/GaN MQW device serves as LED and photodiode simultaneously, an identical fabrication process is adopted to achieve an identical device architecture. The top layer is defined by photolithography and then etched down to n-GaN to form an isolation mesa by inductively-coupled plasma reactive ion etching (ICP-RIE) using $Cl_2$ and $BCl_3$ hybrid gases, and a ~10-μm-wide trench is formed to isolate the two p-regions. Both p- and n-type contacts are then formed by Ni/Au (20 nm/180 nm) evaporation followed by rapid thermal annealing in a $N_2$ atmosphere. The waveguides are then patterned and etched by ICP-RIE. The top device structures are protected and the silicon substrate is then patterned by rear-side alignment photolithography. Deep RIE is performed to remove the silicon substrate and back wafer etching of the suspended membrane is performed to obtain a membrane-type in-plane VLC system.

Figure 1(a) shows an optical microscope image of the fabricated in-plane VLC system, consisting of two suspended p-n junction InGaN/GaN MQW devices and suspended waveguides. These integrated devices are fabricated on a suspended membrane, except for the contact pads, to which four probes are applied for device characterization. Figure 1(b) shows a scanning electron microscope (SEM) image of the in-plane VLC system. Two suspended p-n junction InGaN/GaN MQW devices, in which the gaps between the mesas and the n-electrodes are 10 μm, are connected to 70-μm-diameter p-electrodes fabricated on the isolation mesa using three 80-μm-long and 10-μm-wide suspended waveguides, which are spaced in parallel. The suspended p-n junction InGaN/GaN MQW devices serve as transmitters and receivers simultaneously, and the suspended waveguides are used as in-plane channels to achieve full duplex communication using visible light.

Figure 2 shows a schematic of the experimental setup. The integrated setup includes an Agilent B1500A semiconductor device analyzer, an Agilent 33522A arbitrary waveform generator, a Cascade PM5 probe station and an Agilent DSO9254A digital storage oscilloscope. The probe station observation system can image the sample simultaneously. When the probes are applied to the sample, the devices can be driven by either the semiconductor device analyzer or the signal generator, and the measured data can be characterized using the oscilloscope or other analyzers for integrated device evaluation.

When the suspended p-n junction InGaN/GaN MQW device operates in LED mode, it emits light in a pancake-shaped pattern in all directions within the plane, and there are thus some escape cones that are parallel to the wafer surface [15]. The light emission can thus couple into the suspended waveguides, leading to in-plane propagation [16–21]. Light coupling between the two suspended p-n junction InGaN/GaN MQW devices is achieved via the suspended waveguides. One device at the end of the in-plane VLC system emits light that is coupled into the suspended waveguide, and another device at the other end of the system absorbs that guided light and completes the photon-electron conversion. In association with the semiconductor device analyzer and the probe station, the light coupling behavior of the system is characterized via four-terminal measurements. Figure 3(a) shows log-scaled current-voltage (*I-V*) plots for device B at various forward current levels for device A. The measured current at device B is the sum of the driven current caused by the bias voltage and the induced photocurrent caused by the light emitted by device A. The light emission intensity is enhanced by increasing the forward current of device A; as the forward current of device A increases, the emitted light becomes stronger. Device B can thus absorb increased light power, leading to an increased photocurrent. This induced photocurrent clearly influences the current measured at device B. The induced photocurrents are negative when the device B bias voltage is lower than the turn-on voltage. Therefore, only positive currents are observed for device B when the forward current of device A is in the 0.2 mA to 1 mA range. The device turn-on voltage is approximately 2.5 V at the 0 mA forward current for device A, and increases with increasing forward current through device A.

Linearly-scaled *I-V* plots in the −4 V to 2.5 V range are shown in Fig. 3(b). Device B detects the light guided by the suspended waveguides, leading to in-plane photon-electron conversion inside the p-n junction. The measured photocurrent thus increases with increasing device B bias voltage. To investigate the induced photocurrent further, we use measured current values for device B at a forward current of 0 mA for device A to subtract other measured current values. As shown in Fig. 3(c), below the turn-on voltage for device B, the induced photocurrent gradually increases with increasing bias voltage. Around the turn-on voltage, the induced photocurrent increases sharply before reaching a relatively stable region. The induced photocurrent also clearly increases with increasing forward current through device A. Figure 3(d) shows the induced photocurrent in device A at different forward current levels for device B; similar results are observed. These experimental results confirm that the p-n junction InGaN/GaN MQW devices can detect the light guided by the suspended waveguides when the devices operate in either LED mode or photodiode mode, i.e., simultaneous light emission and photodetection is realized.

The two p-n junction InGaN/GaN MQW devices are driven directly by the arbitrary waveform generator to modulate the emitted light. When device B is turned off, device A is driven directly to modulate the emitted light at 23 MHz. Figure 4(a) shows the light emission image. Because a Ni/Au (20 nm/180 nm) metallization stack is used as the p-electrode, the electrode region is not transparent and the thick metal stack suppresses light emission from the top escape cone, leading to a dark p-electrode region. The light that is emitted from the neighboring region is attributed to the current spreading effect [22]. Figure 4(b) shows the light image when device A is driven at 23 MHz and device B is driven at 2 MHz simultaneously, where the offset voltages are 3.0 V and the peak-to-peak modulation voltages are 4.1 V and 6.5 V, respectively. The peak-to-peak modulation voltage decreases with increasing frequency, leading to a difference between the illumination levels from the two devices. The two p-n junction InGaN/GaN MQW devices serve as both transmitters and receivers simultaneously and, together with the suspended waveguides, form an in-plane VLC system.

In the in-plane VLC system, the emitted light at one system communication port is delivered into the waveguide, and the photon-electron conversion processes are completed at the other port, thus realizing in-plane communication using visible light [23]. Both the transmitted and received electric signals are characterized by the oscilloscope without amplification processes. When device B is driven at 2 MHz and device A is driven at 23 MHz, the signals at device B are divided into two channels, where one is used to produce the received signals with a high-pass filter and the other is used to characterize the transmitted signals without the filter. Figure 5(a) shows the signals for the in-plane VLC system, where device B emits modulated light at 2 MHz and detects the modulated light from device A at 23 MHz. Device A delivers a 23 MHz sinewave signal, and both the 2 MHz and 23 MHz sinewave signals are measured at device B, thus experimentally indicating that the p-n junction InGaN/GaN device can detect modulated light while it also emits modulated light. The oscilloscope trace in Fig. 5(b) shows that the two channels at device B sense the modulated light at 23 MHz when device B is turned off, i.e., illustrating photodiode mode operation. Figure 5(c) shows the measured eye diagrams at 500 kbps for the in-plane VLC system using the probe station, in which open eyes are clearly shown. Additionally, when both device A and device B simultaneously emit light modulated with the 500 kHz square wave signals and with fill factors of 0.5 and 0.25, respectively, figure 5(d) shows that device A can transform both the transmitted and received signal traces into a superposition of signals, leading to an in-plane full duplex VLC system that operates at the same frequency. These results experimentally demonstrate that full duplex communication can be achieved in the same channel using the proposed in-plane VLC system. Self-interference cancellation and other signal processing methods can be used to extract the signals.

In conclusion, suspended p-n junction InGaN/GaN MQW devices and suspended waveguides are fabricated on a single GaN-on-silicon platform. When operated in either LED mode or photodiode mode, the suspended p-n junction InGaN/GaN MQW devices can produce light-induced currents, i.e., simultaneous light emission and photodetection is realized. Using the two suspended p-n junction InGaN/GaN MQW devices and the suspended waveguides, the full in-plane VLC system is constructed. In-plane light coupling between the two suspended p-n junction InGaN/GaN MQW devices is realized using the suspended waveguides. Full duplex communications are demonstrated experimentally using visible light, thus paving the way towards an in-plane full duplex VLC system operating at the same frequency on a single chip.

**Acknowledgements**

This work is jointly supported by the National Natural Science Foundation of China (grant nos. 61322112 and 61531166004), and Research Projects (nos. 2014CB360507, RLD201204, and BJ211026).

**References**

1. H. L. Minh, D. O'Brien, G. Faulkner, L. Zeng, K. Lee, D. Jung, Y. J. Oh, and E. T. Won, "100-Mb/s NRZ Visible Light Communications Using a Postequalized White LED," IEEE Photon. Technol. Lett. **21**, 1063-1065(2009).

2. J. Vučić, C. Kottke, S. Nerreter, K. D. Langer, and J. W. Walewski, "513 Mbit/s visible light communications link based on DMT-modulation of a white LED," J. Lightw. Technol. **28**, 3512-3518(2010).

3. R. Zhang, H. Claussen, H. Haas, and L. Hanzo, "Energy efficient visible light communications relying on amorphous cells," IEEE J. Sel. Area. Comm. **34**, 1-28(2016).

4. D. O'Brien, G. Parry, and P. Stavrinou, "Optical hotspots speed up wireless communication," Nature Photon. **1**, 245-247 (2007)

5. P. Dietz, W. Yerazunis, and D. Leigh. Very low-cost sensing and communication using bidirectional leds. In UbiComp 2003: Ubiquitous Computing, pages 175-191. Springer, 2003.

6. M. D. Brubaker, P. T. Blanchard, J. B. Schlager, A. W. Sanders, A. Roshko, S. M. Du, J. M. Gray, V. M. Bright, N. A. Sanford, and K. A. Bertness, "On-Chip Optical Interconnects Made with Gallium Nitride Nanowires." Nano Lett. **13**, 374-377 (2013).

7. M. Tchernycheva, A. Messanvi, A. de Luna Bugallo, G. Jacopin, P. Lavenus, L. Rigutti, H. Zhang, Y. Halioua, F. H. Julien, J. Eymery, and C. Durand, "Integrated Photonic Platform Based on InGaN/GaN Nanowire Emitters and Detectors." Nano Lett. **14**, 3515-3520(2014).

8. X. Li, G. Y. Zhu, X. M.Gao, D.Bai, X. M. Huang, X. Cao, H. B. Zhu, K.Hane, and Y. J. Wang, "Suspended p-n Junction InGaN/GaN Multiple-Quantum-Well Device With Selectable Functionality," IEEE Photon. J. **7**, 2701407(2015).

9. S. Schmid, G. Corbellini, S. Mangold, and T. R. Gross, "LED-to-LED visible light communication networks," in Proceedings of the fourteenth ACM international symposium on Mobile ad hoc networking and computing. ACM, 2013, pp. 1-10.

10. S. Hong, J. Brand, J. II Choi, M. Jain, J. Mehlman, S. Katti, and P. Levis, "Applications of self-interference cancellation in 5G and beyond," IEEE Commun. Mag. **52**, 114-121(2014).

11. X. Li, Z. Shi, G. Zhu, M. Zhang, H. Zhu and, Y. Wang, "High efficiency membrane light emitting diode fabricated by back wafer thinning technique," Appl. Phys. Lett. **105**, 031109(2014).

12. J. Yuan, W. Cai, X. Gao, G. Zhu, D. Bai, H. Zhu, and Y. Wang, "Monolithic integration of a suspended light-emitting diode with a Y-branch structure," Appl. Phys. Express **9**, 032202 (2016).

13. W. Cai, X. M Gao, W. Yuan, Y. C. Yang, J. L. Yuan, H. B. Zhu and Y.J. Wang, "Integrated p-n junction InGaN/GaN multiple-quantum-well devices with diverse functionalities," Appl. Phys. Express **9**, 052204(2016).

14. Y. J. Wang, G. X. Zhu, W. Cai, X. M. Gao, Y. C. Yang, J. L. Yuan, Z. Shi, and H. B. Zhu, "On-chip photonic system using suspended p-n junction InGaN/GaN multiple quantum wells device and multiple waveguides," Appl. Phys. Lett., **108**, 162102 (2016).

15. D. Bai, T. Wu, X. Li, X. Gao, Y. Xu, Z. Cao, H. Zhu, and Y. Wang, "Suspended GaN-based nanostructure for integrated optics," Appl. Phys. B **122**, 1-7(2016).

16. Y. Zhang, L. McKnight, E. Engin, I. M. Watson, M. J. Cryan, E. Gu, M. G. Thompson, S. Calvez, J. L. O'Brien, and M. D. Dawson, "GaN directional couplers for integrated quantum photonics," Appl. Phys. Lett. **99**, 161119(2011).

17. A. W. Bruch, C. Xiong, B. Leung, M. Poot, J. Han, and H. X. Tang, "Broadband nanophotonic waveguides and resonators based on epitaxial GaN thin films," Appl. Phys. Lett. **107**, 141113(2015).

18. A. Stolz, L. Considine, E. Dogheche, D. Decoster, and D. Pavlidis, "Prospective for Gallium Nitride-based optical waveguide modulators," IEICE Trans Electron. **E95-C**, 1363-1368(2012).

19. T. Sekiya, T. Sasaki, K. Hane, "Design, fabrication, and optical characteristics of freestanding GaN waveguides on silicon substrate," J. Vac. Sci. Technol. B. **33**, 031207(2015).

20. N. VicoTriviño, U. Dharanipathy, J.-F.Carlin, Z. Diao, R. Houdre, and N. Grandjean, "Integrated photonics on silicon with wide bandgap GaN semiconductor," Appl. Phys. Lett.**102**, 081120 (2013).

21. I. Roland, Y. Zeng, Z. Han, X. Checoury, C. Blin, M. El Kurdi, A. Ghrib, S. Sauvage, B. Gayral, C. Brimont, T.Guillet, F. Semond, and P. Boucaud, "Near-infrared gallium nitride two-dimensional photonic crystal platform on silicon,"Appl. Phys.Lett.**105**, 011104 (2014).

22. Y.-C. Lee, F.-S.Hwu, M.-C.Yang, and C.-Y. Lin, "Experimental and Numerical Analysis of p-Electrode Patterns on the Lateral GaN-Based LEDs," J. Lightwave Technol. **32**, 2643-2648 (2014)

23. W. Cai, Y. C. Yang, X. M. Gao, J. L. Yuan, W. Yuan, H. B. Zhu, and Yongjin Wang, "On-chip integration of suspended InGaN/GaN multiple-quantum-well devices with versatile functionalities," Opt. Express **24**, 6004-6010 (2016).

**Figure Captions**

**FIG. 1.** (a) Optical microscope image of fabricated in-plane VLC system. (b) SEM image of the integrated devices.

**FIG. 2.** Schematic diagram of the experimental setup.

**FIG. 3.** (a) Log-scaled *I-V* plots for device B at various forward current levels for device A. (b) Linearly-scaled *I-V* plots for device B at various forward current levels for device A. (c) Log-scaled photocurrents for device B at various forward current levels for device A. (d) Log-scaled photocurrents for device A at various forward current levels for device B.

**FIG. 4.** (a) Light emission image when device A is driven at 23 MHz and device B is turned off. (b) Light emission image when device A is driven at 23 MHz and device B is driven at 2 MHz, where the offset voltages are 3.0 V and the peak-to-peak modulation voltages are 4.1 V and 6.5 V, respectively.

**FIG. 5.** (a) Measured signals for in-plane VLC system when device A is driven at 23 MHz and device B is driven at 2 MHz. (b) Measured signals for in-plane VLC system when device A is driven at 23 MHz and device B is turned off. (c) Eye diagrams measured at 500 kbps. (d) Superposition of signals at device A.

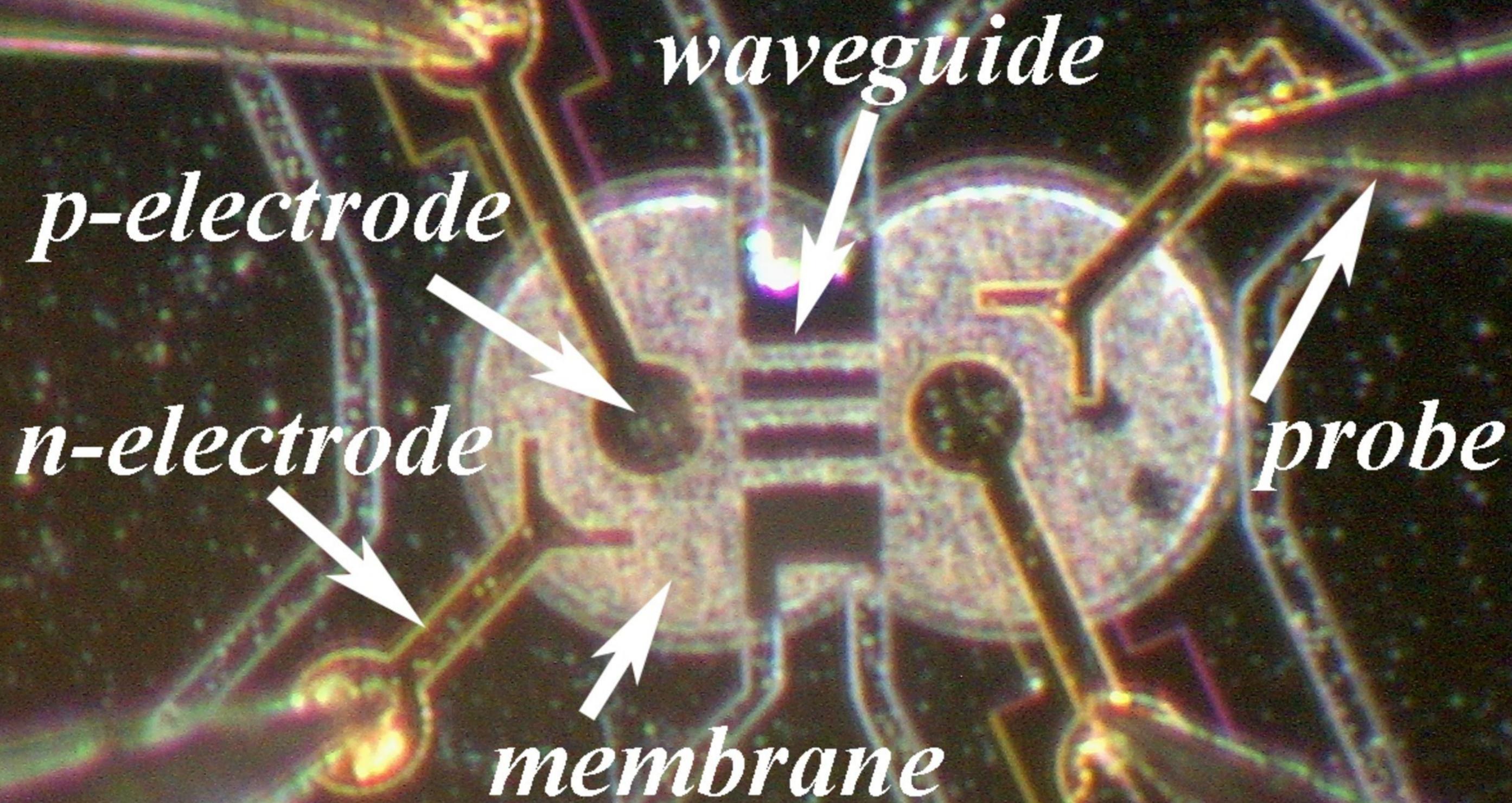

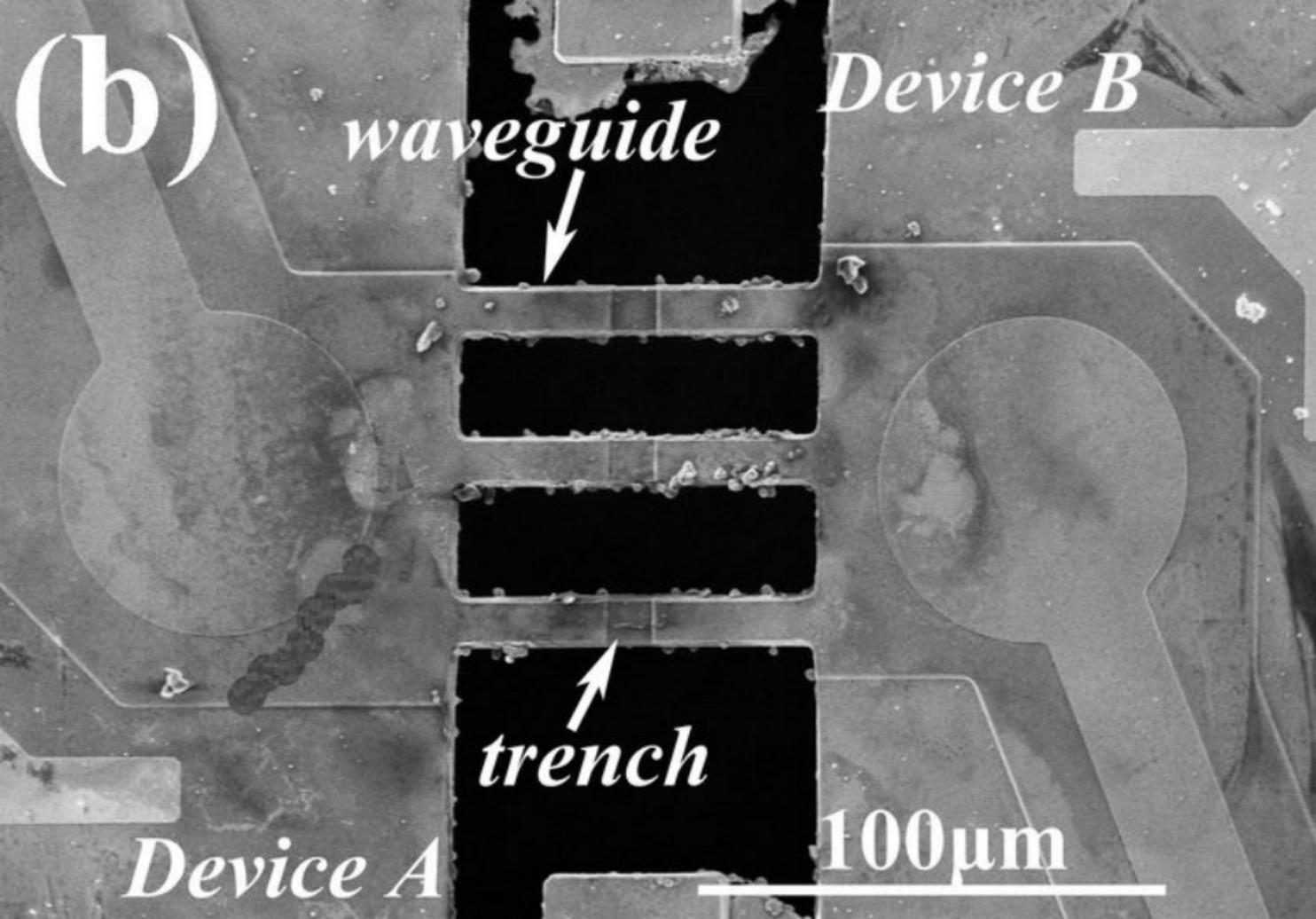

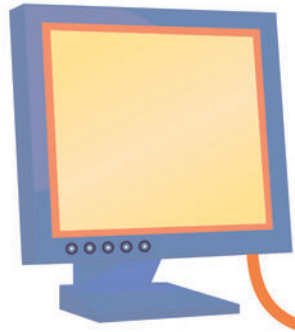
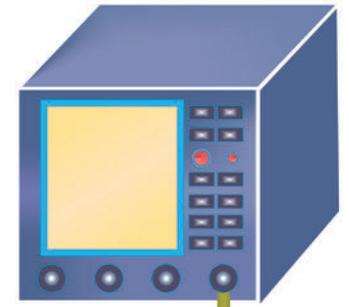
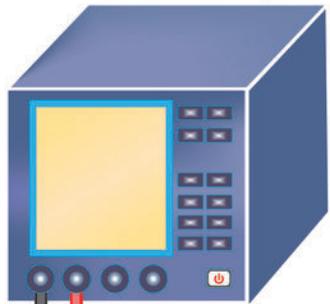
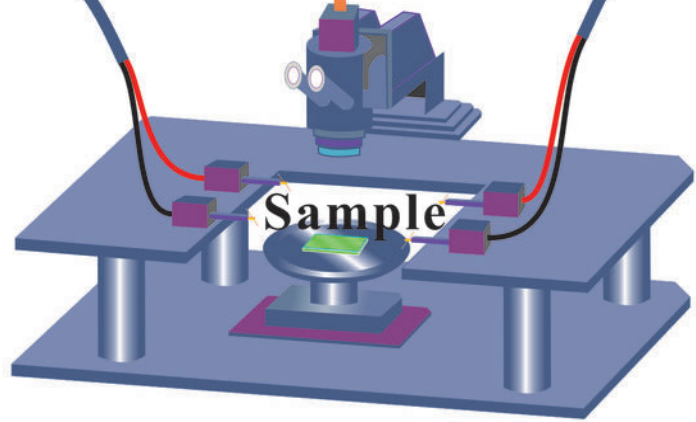
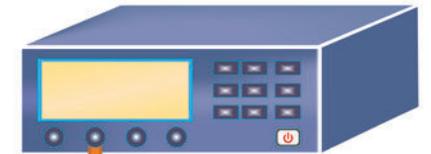
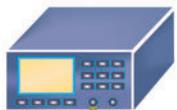
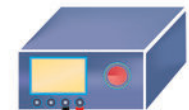

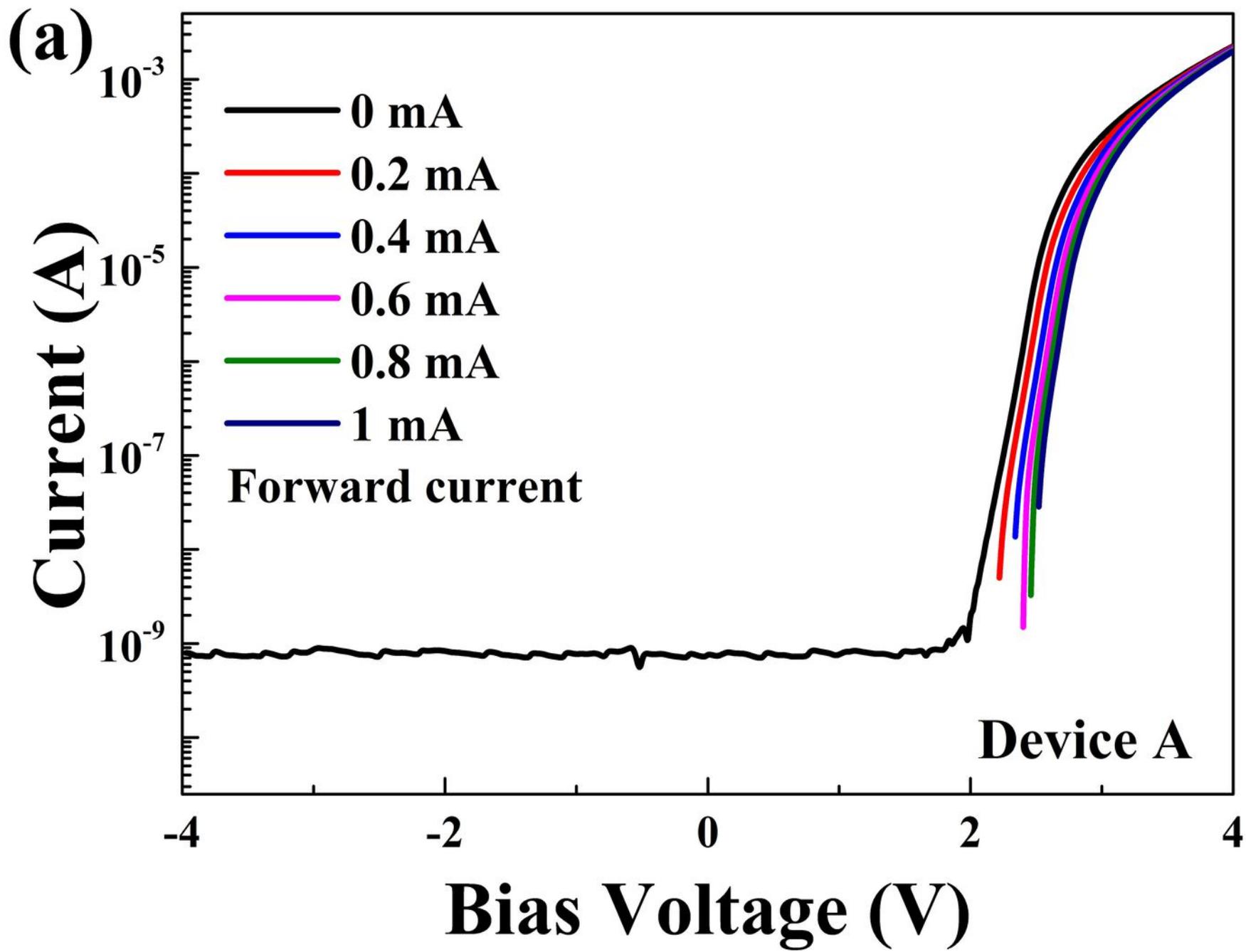

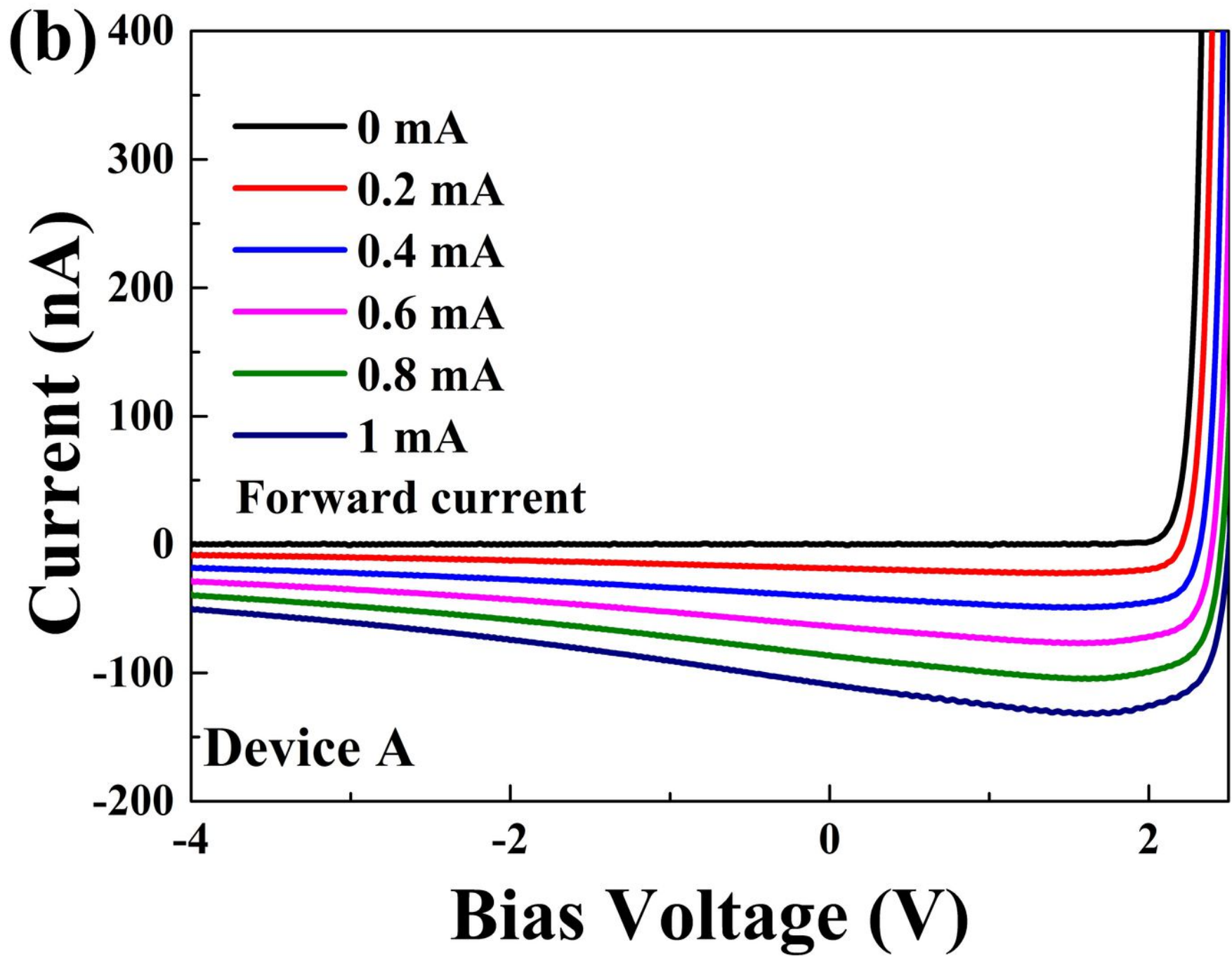

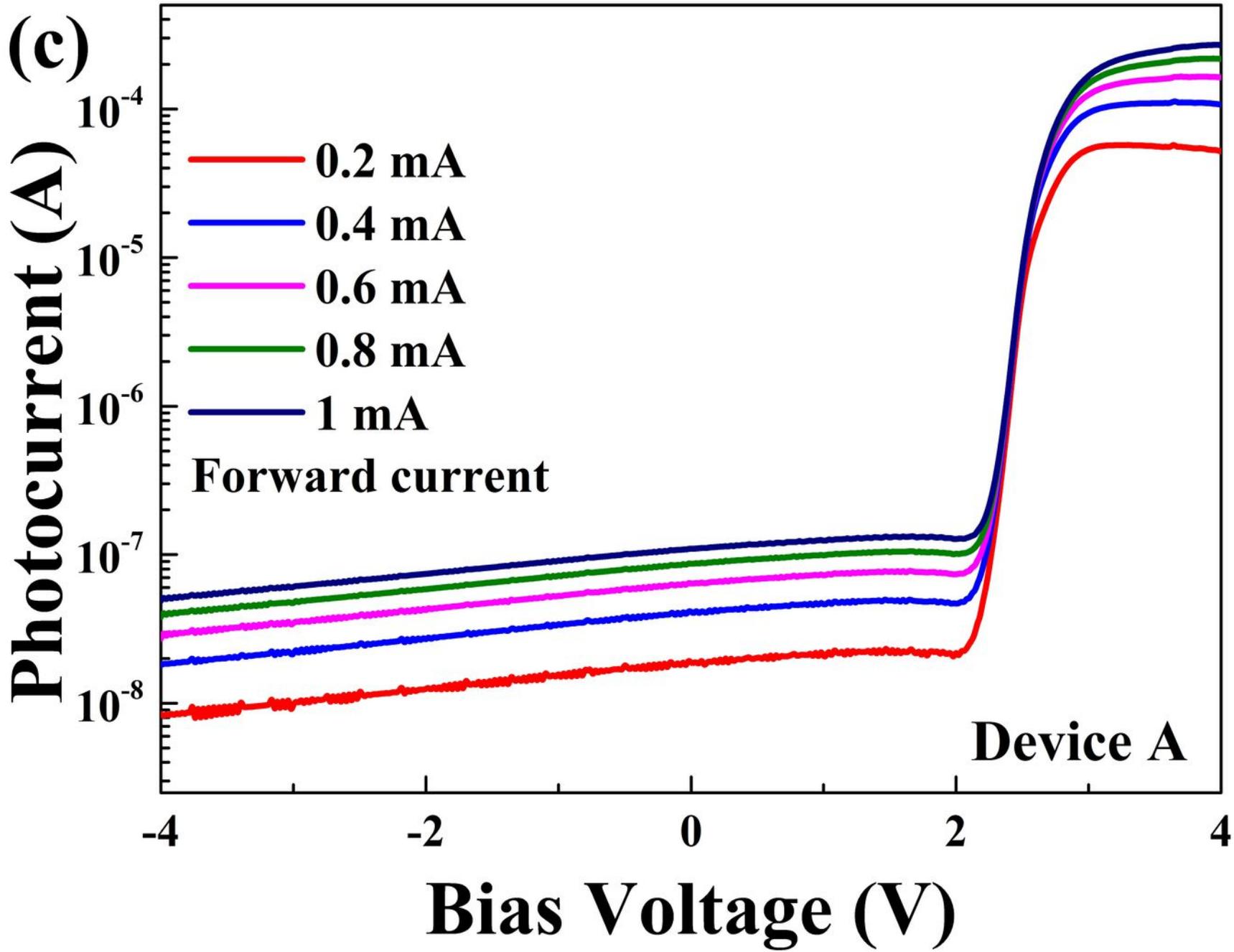

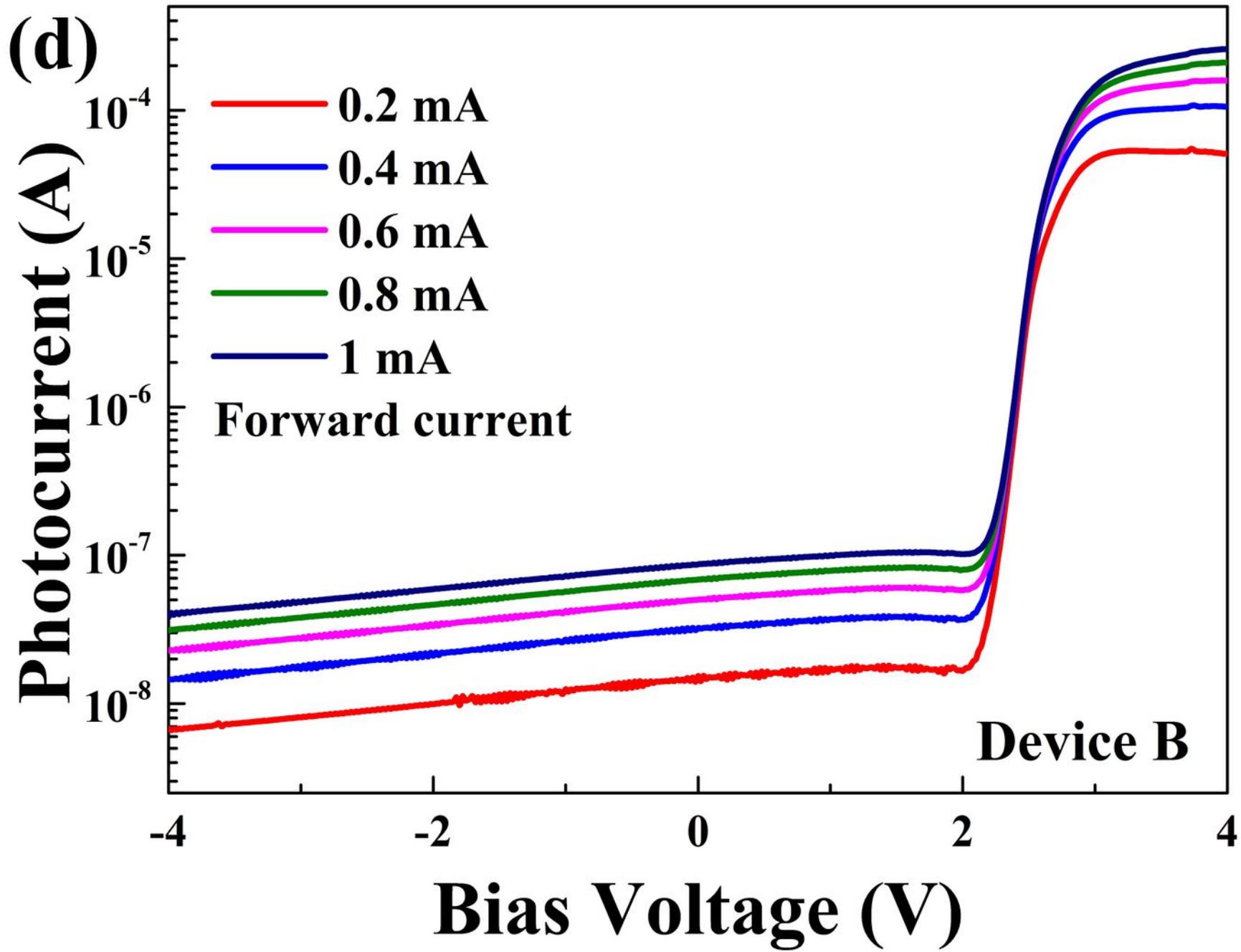

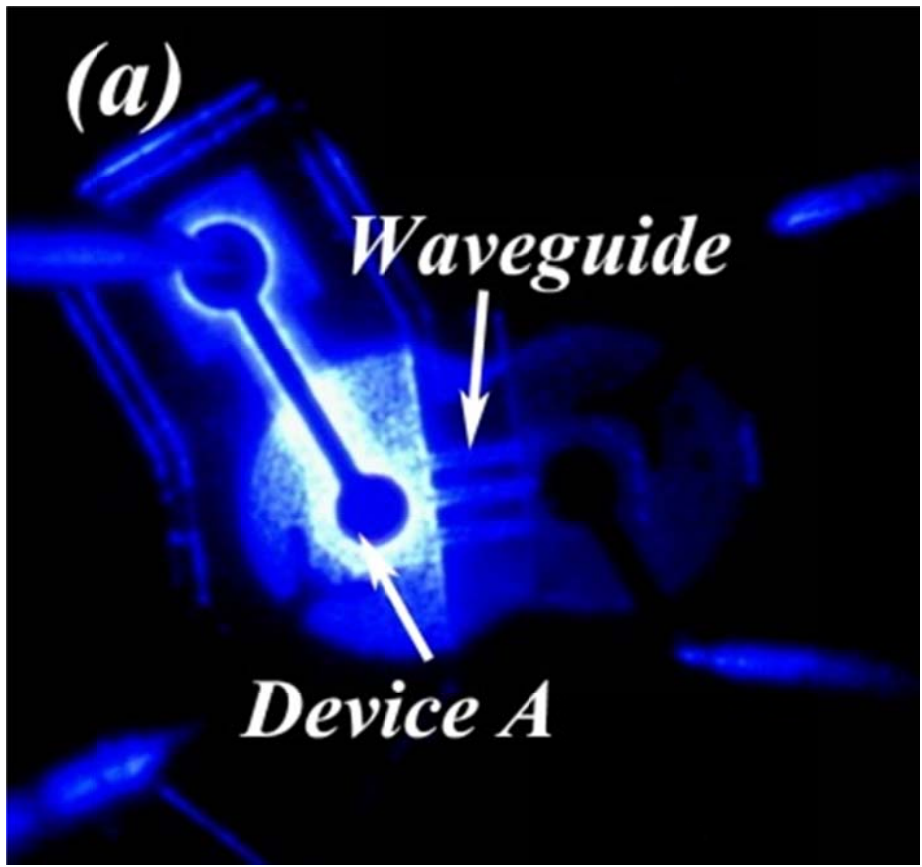
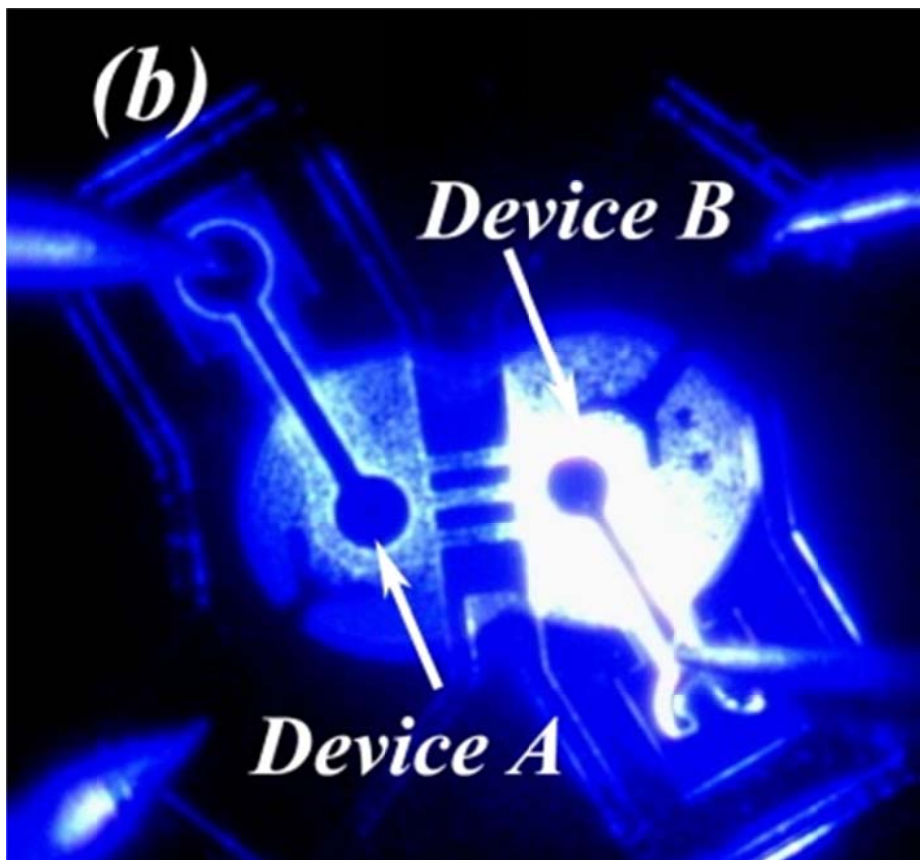

**Figure 4**
*Yongjin Wang et al.*

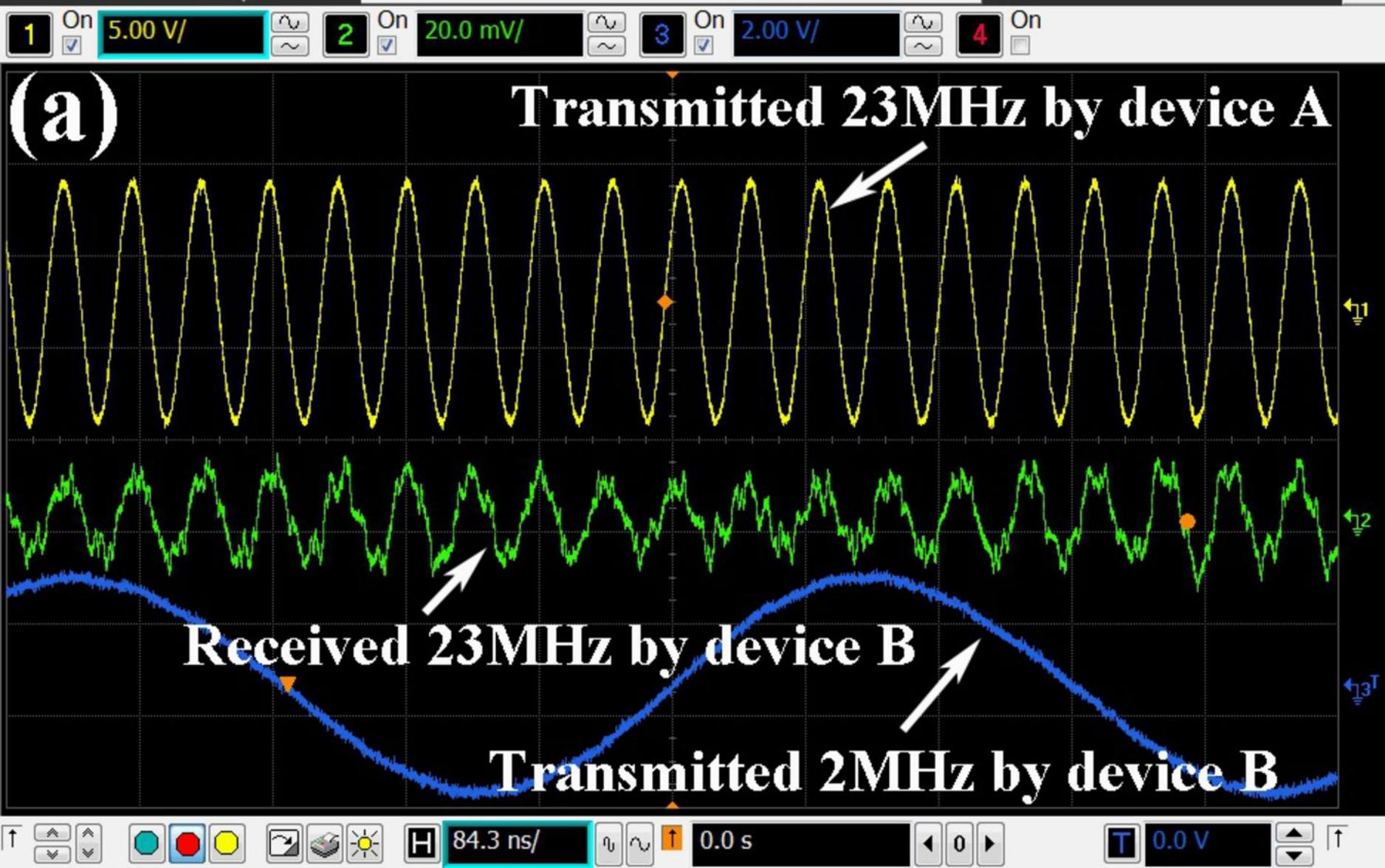

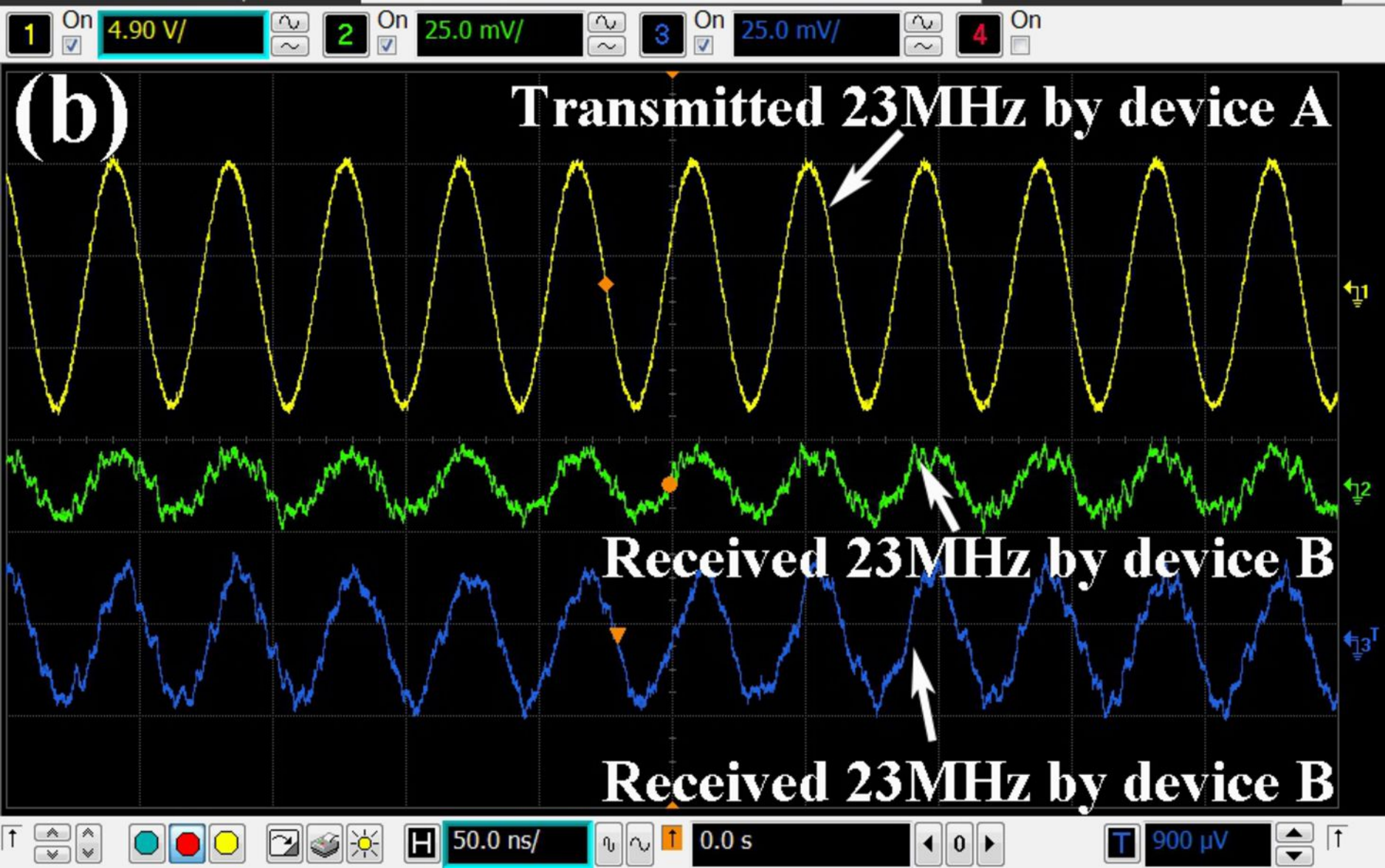

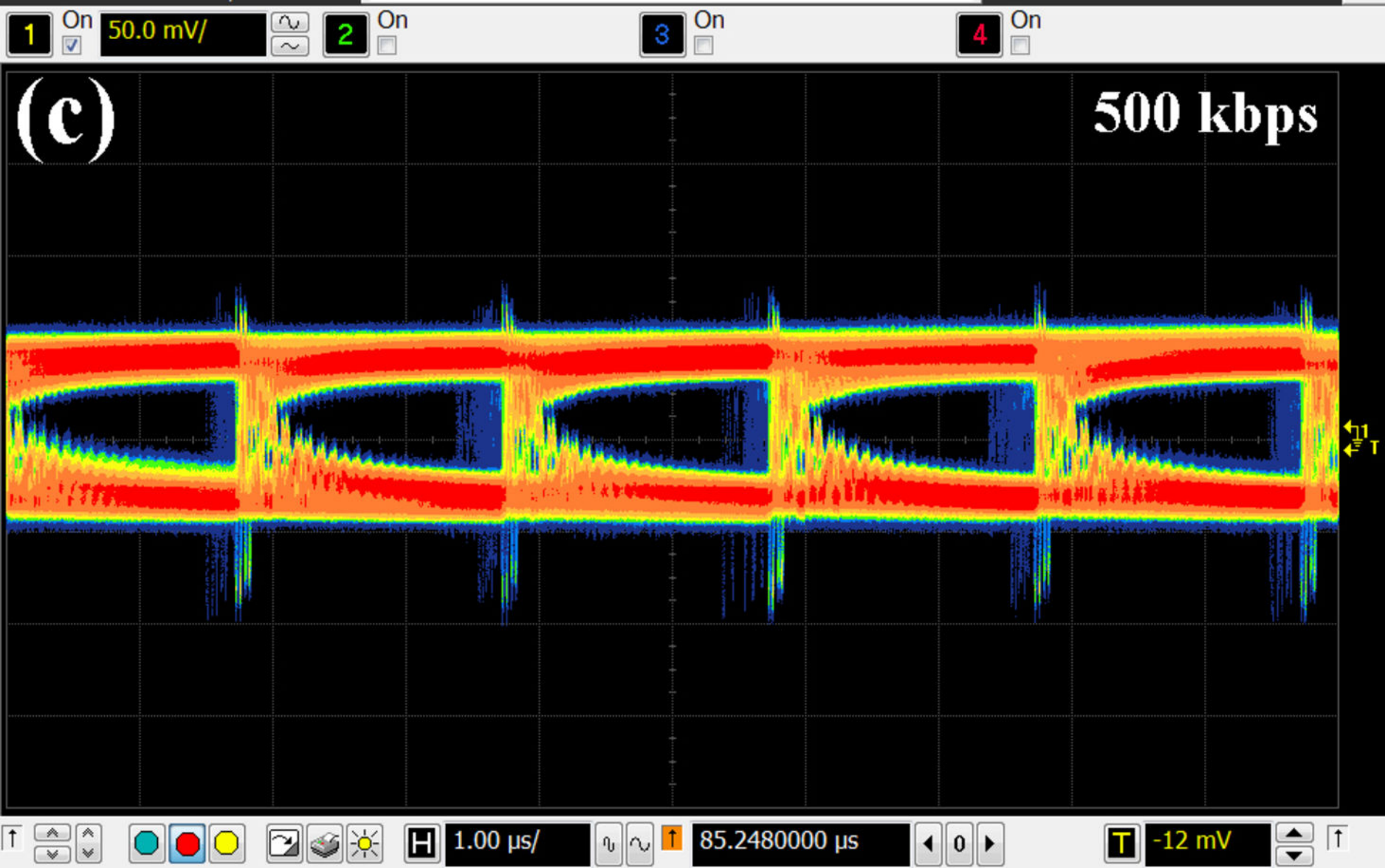

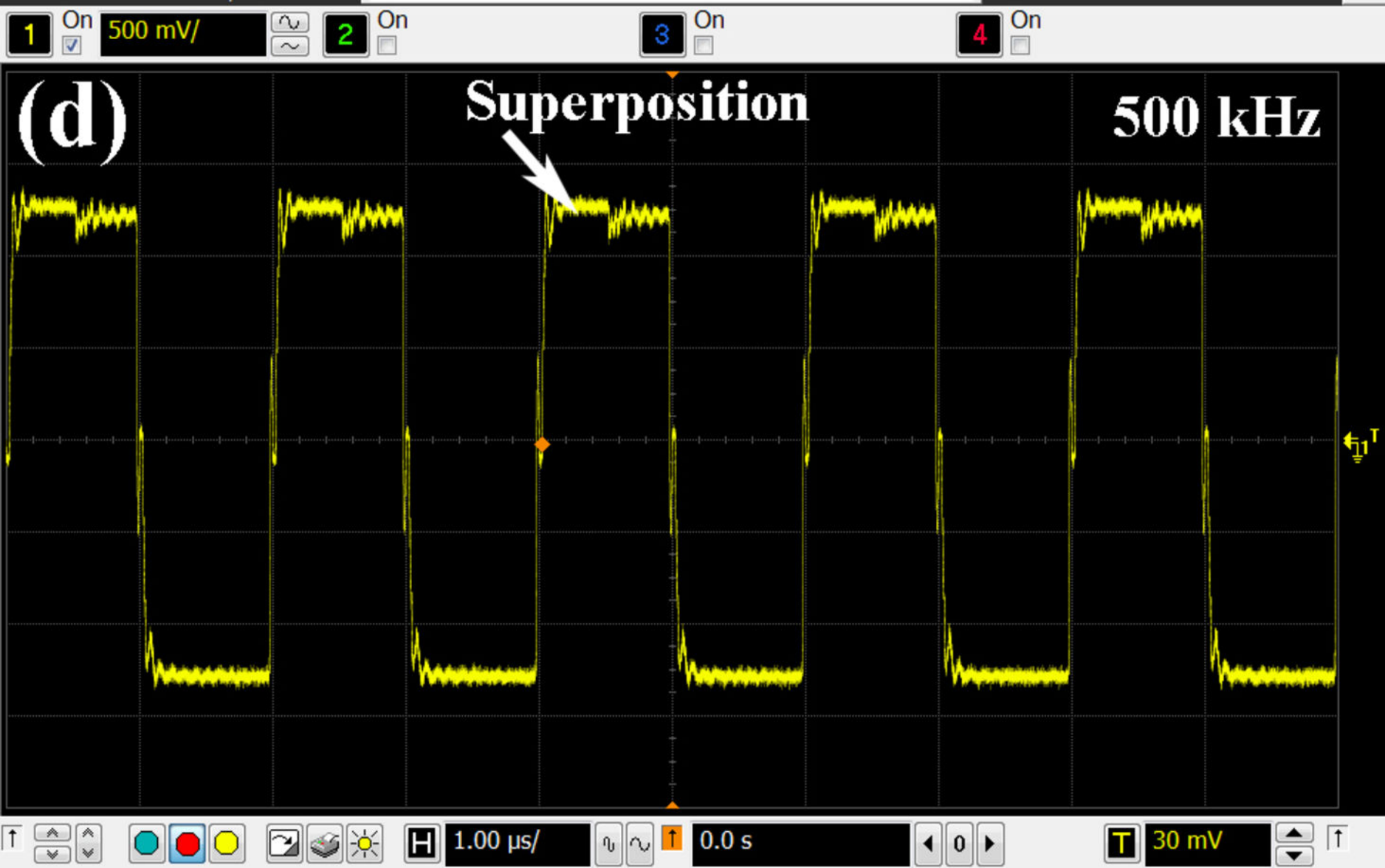